\documentclass[amsmath,amssymb,prl,superscriptaddress,showpacs,longbibliography,twocolumn]{revtex4-1}
\usepackage{graphicx}
\usepackage{dcolumn}
\usepackage[colorlinks=true,linkcolor=blue,citecolor=blue,urlcolor=blue]{hyperref}
\usepackage{stmaryrd}
\usepackage{esvect}
\usepackage[usenames,dvipsnames]{xcolor}
\usepackage{soul}
\usepackage{upgreek}

\newcommand{\ud}{\mathrm{d}}

\newcommand{\ImTr}{\mathrm{Im}\,\mathrm{Tr}}

\newcommand{\EF}{E_{\text{F}}}
\newcommand{\MC}[1]{\mathcal{#1}}
\newcommand{\MR}[1]{\mathrm{#1}}
\newcommand{\VEC}[1]{\mathbf{#1}}

\usepackage{siunitx}
\DeclareSIUnit\mub{\mu_\text{B}}
\usepackage{mathtools}

\DeclarePairedDelimiterX\braket[1]{\langle}{\rangle}{#1}

\begin{document}

\title{Reply to `Comment on ``Proper and improper chiral magnetic interactions'' '}

\author{Manuel dos Santos Dias}\email{m.dos.santos.dias@fz-juelich.de}
\affiliation{Peter Gr\"{u}nberg Institut and Institute for Advanced Simulation, Forschungszentrum J\"{u}lich \& JARA, 52425 J\"{u}lich, Germany}
\affiliation{Faculty of Physics, University of Duisburg-Essen and CENIDE, 47053 Duisburg, Germany}
\author{Sascha Brinker}
\affiliation{Peter Gr\"{u}nberg Institut and Institute for Advanced Simulation, Forschungszentrum J\"{u}lich \& JARA, 52425 J\"{u}lich, Germany}
\author{Andr\'as L\'aszl\'offy}
\affiliation{Wigner Research Centre for Physics, P.O. Box 49, H-1525 Budapest, Hungary}
\affiliation{Department of Theoretical Physics, Budapest University of Technology and Economics, Budafoki \'ut 8, H-1111 Budapest, Hungary}
\author{Bendeg\'uz Ny\'ari}
\affiliation{Department of Theoretical Physics, Budapest University of Technology and Economics, Budafoki \'ut 8, H-1111 Budapest, Hungary}
\author{Stefan Bl\"ugel}
\affiliation{Peter Gr\"{u}nberg Institut and Institute for Advanced Simulation, Forschungszentrum J\"{u}lich \& JARA, 52425 J\"{u}lich, Germany}
\author{L\'aszl\'o Szunyogh}
\affiliation{Department of Theoretical Physics, Budapest University of Technology and Economics, Budafoki \'ut 8, H-1111 Budapest, Hungary}
\affiliation{MTA-BME Condensed Matter Research Group, Budapest University of Technology and Economics, Budafoki \'ut 8, H-1111 Budapest, Hungary}
\author{Samir Lounis}\email{s.lounis@fz-juelich.de}
\affiliation{Peter Gr\"{u}nberg Institut and Institute for Advanced Simulation, Forschungszentrum J\"{u}lich \& JARA, 52425 J\"{u}lich, Germany}
\affiliation{Faculty of Physics, University of Duisburg-Essen and CENIDE, 47053 Duisburg, Germany}

\date{\today}

\begin{abstract}
In our previous Letter [Phys.\ Rev.\ B \textbf{103}, L140408 (2021)], we presented a discussion of the fundamental physical properties of the interactions parameterizing atomistic spin models in connection to first-principles approaches that enable their calculation for a given material.
This explained how some of those approaches can apparently lead to magnetic interactions that do not comply with the expected physical properties, such as Dzyaloshinskii-Moriya interactions which are non-chiral and independent of the spin-orbit interaction, and which we consequently termed `improper'.
In the preceding Comment [Phys.\ Rev.\ B \textbf{105}, 026401], the authors present arguments based on the distinction between global and local approaches to the mapping of the magnetic energy using first-principles calculations to support their proposed non-chiral Dzyaloshinskii-Moriya interactions and their dismissal of our distinction between `proper' and `improper' magnetic interactions.
In this Reply, we identify the missing step in the local approach to the mapping and explain how ignoring this step leads to the identification of magnetic interactions which do not comply with established physical principles and that we have previously termed `improper'.
\end{abstract}

\maketitle

\emph{Context.}
The main issue being addressed in our previous Letter~\cite{dosSantosDias2021} and the subsequent Comment~\cite{Cardias2022} concerns the mapping of density functional theory (DFT) calculations to atomistic spin models and the corresponding interpretation.
This is an intricate and somewhat subtle issue, so it is important to carefully define the quantities involved and to give them informative names, and to relate the magnetic interactions to fundamental microscopic models, as we stressed in our Letter.
In the Comment, the authors first present an overview of mapping approaches, highlighting the prominent role played by the Liechtenstein-Katsnelson-Antropov-Gubanov (LKAG) approach~\cite{Liechtenstein1984,Liechtenstein1987} and how its extension to noncollinear magnetic states leads to what is therein termed the nonrelativistic or non-chiral Dzyaloshinskii-Moriya interaction (DMI).
In contrast to many other known magnetic interactions which are achiral, the original DMI~\cite{Dzyaloshinskii1957,Moriya1960} which is familiar to the magnetism community is chiral and originates from spin-orbit coupling (SOC).
This is followed by a discussion of the content of the corresponding LKAG-type formulae based on form of the Kohn-Sham hamiltonian for noncollinear magnetic systems, which recaps arguments given in Refs.~\cite{Cardias2020,Cardias2020a}.
The parametrization of atomistic spin models using the noncollinear LKAG approach is then compared to our parametrization given in Ref.~\cite{dosSantosDias2021}, which is based on the spin cluster expansion, listing the corresponding advantages and disadvantages.
The essential differences between the noncollinear LKAG approach and the spin cluster expansion are finally discussed in terms of local vs.\ global approaches to the mapping of the magnetic energy, pointing to the recent Ref.~\cite{Streib2021}, with a final remark arguing in favor of the name `nonrelativistic or non-chiral DMI'.
Overall, we find the Comment very constructive and a valuable contribution to the scientific discussion of the important problem of parameterizing the magnetic energy of real materials using DFT approaches.
However, it is our view that the authors have overlooked important aspects of the mapping between first-principles calculations and atomistic spin models using LKAG-type approaches which are central to the interpretation of the calculations in terms of magnetic interactions.
These aspects were discussed in Ref.~\cite{dosSantosDias2021} and here we expand upon them in connection to the preceding Comment.

\emph{Mapping from DFT to a spin model.}
Within DFT, the central quantities describing the energetics of a given material are the electronic charge and spin densities, $n(\VEC{r})$ and $\VEC{m}(\VEC{r})$, respectively, while an atomistic spin model is specified in terms of classical or quantum spins, $\VEC{S}_i$.
The mapping that we are discussing is then represented by
\begin{equation}\label{eq:mapping}
    E[n(\VEC{r}),\VEC{m}(\VEC{r});\VEC{S}_1,\ldots,\VEC{S}_N] \;\longrightarrow\; \MC{E}(\VEC{S}_1,\ldots,\VEC{S}_N) \;.
\end{equation}
The mapping indicated by the arrow comprises three aspects: (i) how to define the total energy functional for a target spin configuration, $E[n(\VEC{r}),\VEC{m}(\VEC{r});\VEC{S}_1,\ldots,\VEC{S}_N]$, (ii) how to find the parameters for the atomistic spin model $\MC{E}$ using the DFT total energy functional $E$, and (iii) under what conditions is the mapping meaningful.
We will focus on the relation between (ii) and (iii).

\emph{LKAG-type approaches to the mapping.}
The LKAG approach to point (ii) consists in expanding the total energy of the magnetic material with respect to small deviations of the orientations of the magnetic moments from a chosen reference magnetic state.
This was the ferromagnetic state in the original publications~\cite{Liechtenstein1984,Liechtenstein1987} and was later generalized to noncollinear reference magnetic states~\cite{Antropov1997,Antropov1999,Lounis2010a,Szilva2013,Secchi2015,Szilva2017,Cardias2020,Cardias2020a}.
Thus the LKAG approach is to be understood as a Taylor expansion of the magnetic energy that gives a local view of the magnetic energy landscape, as also advocated in the Comment and in Ref.~\cite{Streib2021}.

Considering the mapping expressed in Eq.~\eqref{eq:mapping} and deviations of a reference state written as $\VEC{S}_i = \VEC{S}_i^0 + \delta\VEC{S}_i$, we not only need to Taylor expand the DFT total energy,
\begin{align}\label{eq:taylorexp}
    E[\{\VEC{S}_i\}] &= E[\{\VEC{S}_i^0\}] 
    + \sum_i\sum_\mu \frac{\delta E[\{\VEC{S}_i^0\}]}{\delta S_i^\mu}\,\delta S_i^\mu \nonumber\\
    &+ \frac{1}{2} \sum_{i,j} \sum_{\mu,\nu} \frac{\delta^2 E[\{\VEC{S}_i^0\}]}{\delta S_i^\mu\delta S_j^\nu}\,\delta S_i^\mu \delta S_j^\nu + \ldots
\end{align}
where $i,j$ label the spin sites and $\mu,\nu=x,y,z$ the spin components, but also the atomistic spin model $\MC{E}[\{\VEC{S}_i\}]$.
The mapping in Eq.~\eqref{eq:mapping} then becomes a mapping between the matching coefficients of the respective Taylor expansions,
\begin{equation}\label{eq:taylormap}
    \frac{\delta E[\{\VEC{S}_i^0\}]}{\delta S_i^\mu} \longrightarrow \frac{\delta \MC{E}[\{\VEC{S}_i^0\}]}{\delta S_i^\mu} \;,\quad
    \frac{\delta^2 E[\{\VEC{S}_i^0\}]}{\delta S_i^\mu\delta S_j^\nu} \longrightarrow \frac{\delta^2 \MC{E}[\{\VEC{S}_i^0\}]}{\delta S_i^\mu\delta S_j^\nu} \;\ldots
\end{equation}
To establish this mapping, the form of $\MC{E}[\{\VEC{S}_i\}]$ must be specified a priori.
In \cite{dosSantosDias2021} we showed that applying these ideas to a model including isotropic four-spin interactions results in a Taylor expansion that contains DMI-like terms and other anisotropic terms if the reference magnetic state $\{\VEC{S}_i^0\}$ is noncollinear (see e.g.\ Eq.\ 8 in \cite{dosSantosDias2021}).
This is a counterexample against the direct interpretation of the terms of the Taylor expansion of the magnetic energy as magnetic interactions, as in this case the interactions were given a priori and their contributions to the LKAG-type expansion could be traced term by term.

\emph{Interpretation of the Taylor expansion.}
What is discussed in the Comment and in Refs.~\cite{Cardias2020,Cardias2020a,Mankovsky2021} is the microscopic origin of the different coefficients in the Taylor expansion of the magnetic energy for a noncollinear magnetic state, by relating the electronic Green functions to charge and spin currents.
We find this very interesting, noting existing interpretations of the DMI in terms of spin currents~\cite{Kikuchi2016,Freimuth2017a}, and we highlight the strong parallel with the linear response theory of electronic transport in magnetic materials, which has a similar interpretation.
Quantities such as anomalous or topological Hall conductivities are theoretically obtained by perturbing a reference magnetic state with an electric field, and a strong emphasis is placed in their variation with a change of the magnetic state (e.g.\ rotation of the magnetic moments with respect to the lattice or change in their relative alignment), as this offers clues concerning the different underlying physical mechanisms (SOC, emergent magnetic fields, etc.) and is experimentally feasible.
Both the conductivities and the coefficients in the LKAG-type expansion of Eq.~\eqref{eq:taylorexp} are thus expected to depend on and change with the choice of reference magnetic state $\{\VEC{S}_i^0\}$, and the precise nature of these variations is what contains information about the underlying interactions.

\emph{Explicit and implicit constructions of the spin model.}
The connection in Eq.~\eqref{eq:taylormap} is needed to interpret the magnetic interactions obtained from the Taylor expansion of the DFT total energy, which in turn requires the form of the atomistic spin model $\MC{E}[\{\VEC{S}_i\}]$, i.e.\ which magnetic interactions it contains.
This can be established in several ways, using phenomenological arguments based on symmetry or by explicit construction starting from a microscopic electronic model, as mentioned both in our Letter and in the Comment.
These constructions produce magnetic interactions that comply with general requirements such as invariance under time-reversal and compliance with the spatial symmetries of the material.
The microscopic approach has the advantage of also identifying the mechanism behind each type of magnetic interaction.
For instance, Refs.~\cite{Takahashi1977,MacDonald1988,Hoffmann2020} derive systematically the magnetic interactions from a Hubbard model without considering SOC, and find isotropic four-spin interactions but no non-chiral DMI.
The Kramers-Anderson superexchange theory employed by Moriya also required SOC in order to generate the DMI~\cite{Moriya1960}.

As an alternative, we can extract the implicit form of the atomistic spin model directly from Eq.~\eqref{eq:taylorexp}.
Adopting the electronic hamiltonian given in Eq.~1 of the Comment and splitting the spin-independent ($\MC{H}^0$) from the spin-dependent parts while omitting the spatial arguments:
\begin{equation}
    \MC{H}^\MR{el} = \MC{H}^0 + \sum_i\,B_i^\MR{xc}\,\VEC{S}_i\cdot\boldsymbol{\upsigma} \;.
\end{equation}
Here we neglect SOC and make explicit the dependence of the exchange-correlation magnetic fields on the directions of the spin moments, with $B_i^\MR{xc}$ the local magnitude of the fields and $\boldsymbol{\upsigma}$ the vector of Pauli matrices.
The second-order coefficients of the Taylor expansion in Eq.~\eqref{eq:taylorexp} follow from the general LKAG-type expression~\cite{Liechtenstein1984,Liechtenstein1987,Antropov1997,Antropov1999,Udvardi2003,Ebert2009,Lounis2010a,Szilva2013,Secchi2015,Szilva2017,Cardias2020,Cardias2020a,Mankovsky2020,Lounis2020,Streib2021}, here given for a pair of magnetic moments:
\begin{equation}\label{eq:lkag}
    \delta^2 E_{12} = -\frac{1}{\pi}\,\ImTr\int^{\EF}\hspace{-1em}\ud E\; B_1^\MR{xc}\,\delta\VEC{S}_1\cdot\boldsymbol{\upsigma}\,G_{12}\,B_2^\MR{xc}\,\delta\VEC{S}_2\cdot\boldsymbol{\upsigma}\,G_{21} \;.
\end{equation}
The trace is over spin and orbital degrees of freedom and $G(E) = \left(E - \MC{H}^\MR{el}\right)^{-1}$.
We follow the strategy of Ref.~\cite{Brinker2019} in order to extract the dependence of the coefficients on the reference magnetic state:
Introduce $G^0(E) = \left(E - \MC{H}^0\right)^{-1}$, expand the Dyson equation relating $G$ to $G^0$ in the local fields $B_i^\MR{xc}\VEC{S}_i^0$ present in the reference state, and evaluate the spin trace.
The lowest order contribution is found to be
\begin{equation}\label{eq:lkag2}
    \delta^2 E_{12}^{(0)} = -\frac{1}{\pi}\,\ImTr\int^{\EF}\hspace{-1em}\ud E\; B_1^\MR{xc}\,G_{12}^0\,B_2^\MR{xc}\,G_{21}^0\,\big(\delta\VEC{S}_1\cdot\delta\VEC{S}_2\big) \;.
\end{equation}
It has the expected form of the isotropic Heisenberg interaction and is independent of the reference magnetic state.
The next non-vanishing contributions involve the spin directions of the reference state twice, with one example being (note that $\VEC{S}_i^0\cdot\delta\VEC{S}_i = 0$)
\begin{align}\label{eq:lkag4}
    \delta^2 E_{12}^{(2)} &= -\frac{1}{\pi}\,\ImTr\int^{\EF}\hspace{-1em}\ud E\; \left(B_1^\MR{xc}\,G_{12}^0\,B_2^\MR{xc}\,G_{21}^0\right)^2 \nonumber\\
    &\left(\big(\VEC{S}_1^0\cdot\VEC{S}_2^0\big)\big(\delta\VEC{S}_1\cdot\delta\VEC{S}_2\big)
    - \big(\VEC{S}_1^0\times\VEC{S}_2^0\big)\cdot\big(\delta\VEC{S}_1\times\delta\VEC{S}_2\big)\right) \;.
\end{align}
The first term has the form of the isotropic Heisenberg exchange if only the dependence on $\delta\VEC{S}_1\cdot\delta\VEC{S}_2$ is considered, but together with the dependence on $\VEC{S}_1^0\cdot\VEC{S}_2^0$ it resembles the isotropic biquadratic interaction.
Likewise, the second term has the form of the non-chiral DMI advocated by the authors of the Comment if only the dependence on $\delta\VEC{S}_1\times\delta\VEC{S}_2$ is considered, but together with $\VEC{S}_1^0\times\VEC{S}_2^0$ it resembles in fact a rewriting of the isotropic biquadratic interaction, as we discussed in the Letter.
The perturbative expansion can be easily generalized to multiple magnetic moments and to include SOC~\cite{Brinker2019}, and as we just showed it provides the form of the atomistic spin model directly from the LKAG-type expressions.
It also explains why the coefficients in Eq.~\eqref{eq:taylorexp} can vary strongly with a change in the reference magnetic state.

The preceding discussion leads us to conclude that the non-chiral DMI is not a fundamental magnetic interaction, as it cannot be derived from any of those well-established approaches.
It arises from considering the coefficients of the Taylor expansion in Eq.~\eqref{eq:taylorexp} without constructing the associated mapping given by Eq.~\eqref{eq:taylormap}, which obscures the dependence of the coefficients on the choice of the reference magnetic state $\{\VEC{S}_i^0\}$.

\emph{Global vs.\ local mapping approaches to the magnetic energy.}
We agree with the distinction between global and local approaches to the mapping of the magnetic energy discussed at the end of the Comment and in Ref.~\cite{Streib2021}.
While local approaches characterize the energy landscape near a reference magnetic state (in the spirit of LKAG), global approaches parameterize models that reproduce the energies of different magnetic states.
It follows that local and global approaches have different ranges of validity and applicability (our point (iii) in connection to Eq.~\eqref{eq:mapping}).
For instance, local approaches are suitable to compute the magnon spectrum of the magnetic ground state, while global approaches are required to establish the complete magnetic phase diagram in an unbiased way.
As explained in our Letter and is motivated by the preceding discussion, it is our view that the atomistic spin model parametrized through a global approach is more fundamental, as it complies with all the physical requirements that we used to define `proper' magnetic interactions.
The authors of the Comment are correct to point out that the coefficients of the LKAG expansion worked out and computed in Refs.~\cite{Cardias2020,Cardias2020a,Mankovsky2020} do not have to comply with those physical requirements, as we argued that these coefficients should not be directly interpreted as specific magnetic interactions.

\emph{Comparing parameterizations.}
On the section `Comparing Parametrizations' of the Comment, the authors write that the global approach to the magnetic energy used in our Letter does not have a unique mapping onto a multi-spin model.
This is true concerning the calculations for Mn$_3$Sn, where the fitted interaction parameters correspond to effective sublattice interactions that sum over many different types of multi-spin and multi-site interactions.
We remark that we also included interactions among three sublattices and that this was important for a good fit to the angular dependence of the DFT total energies.
However, our global approach based on the spin cluster expansion is unique, systematic and complete for the magnetic trimers, due to the finite magnetic configuration space.
Thus it was quite puzzling that we could not reproduce the very large non-chiral DMI reported in Ref.~\cite{Cardias2020a} for the Cr trimers on Au(111) using our global approach, c.f.\ also previous work in Refs.~\cite{Antal2008,Brinker2020a}.
This disagreement might be settled by comparing different parameterizations to the angular dependence of the total energy for the case of the trimers.

\emph{Conclusions.}
In this Reply we addressed the various points raised by the preceding Comment by focusing on the nature of the mapping between DFT total energy calculations and atomistic spin models, as indicated by Eq.~\eqref{eq:mapping}.
The local approach to this mapping selects a given magnetic reference state and proceeds by Taylor expanding the DFT total energy for small spin deviations around that reference state, which is the spirit of the LKAG approach~\cite{Liechtenstein1984,Liechtenstein1987}.
Eq.~\eqref{eq:taylormap} identifies the missing step in this approach, which demands that the atomistic spin model be subjected to the same Taylor expansion and the expansion coefficients matched to the ones from the DFT total energy.
In our Letter we already provided a simple example of how the Taylor expansion of the DFT total energy can contain coefficients that have the appearance of improper magnetic interactions, such as the non-chiral DMI, a puzzle which is resolved in general by the procedure encoded in Eq.~\eqref{eq:taylormap} and with a concrete example based on the LKAG-type expressions in Eq.~\eqref{eq:lkag4}.

It is our understanding that the authors of the Comment interpret the coefficients of the DFT-based LKAG expansion in Eq.~\eqref{eq:taylorexp} directly as magnetic interactions, according to the form of the combination of spin components involved, and do not separately and independently specific the form of the atomistic spin model according to a set of physical principles, thus skipping the step given in Eq.~\eqref{eq:taylormap}.
The resulting `local' magnetic interactions have the symmetry properties of the magnetic space group of the reference state chosen for the Taylor expansion, in analogy with the well-known properties of electronic transport coefficients in magnetic materials, such as the anomalous or topological Hall conductivities, and similarly can be given microscopic interpretations in terms of charge and spin currents, as also done recently in Ref.~\cite{Mankovsky2021} for three-spin interactions.

Our conclusion is that taking the coefficients of the LKAG expansion to represent directly the magnetic interactions of a given material needs to be reconsidered, and that the terminology of `improper' vs.\ `proper' magnetic interactions introduced in our Letter should be retained given the raised issues with this interpretation.
In particular, we showed that the proposed `non-chiral DMI' cannot be justified either from fundamental microscopic models or even through the analysis of LKAG-type expressions, and so should not be considered as a fundamental magnetic interaction, in contrast to the original chiral one introduced by Dzyaloshinskii and Moriya.

\emph{Acknowledgements.}
We gratefully acknowledge financial support from the DARPA TEE program through grant MIPR (\# HR0011831554) from DOI, from the European Research Council (ERC) under the European Union's Horizon 2020 research and innovation program (Grant No.\ 856538, project ``3D MAGiC'' and ERC-consolidator grant 681405 --- DYNASORE),  from Deutsche For\-schungs\-gemeinschaft (DFG) through SPP 2137 ``Skyrmionics'' (Project BL 444/16 and LO 1659/8-1), Priority Programme SPP 2244 ``2D Materials - Physics of van der Waals Heterostructures'' (project LO 1659/7-1), and the Collaborative Research Centers SFB 1238 (Project C01).
B.N., A.L. and L.S. acknowledge the support provided by the Ministry for Innovation and Technology and by the National Research, Development, and Innovation Office under Project No. K131938 and within the Quantum Information National Laboratory of Hungary.
The authors gratefully acknowledge the computing time granted through JARA-HPC on the supercomputer JURECA at the Forschungszentrum J\"ulich~\cite{jureca}.

\bibliography{references.bib}

\end{document}